\documentclass[amssymb,amsmath,aps,showpacs,floatfix,nofootinbib,showpacs,12pt]{revtex4}
\usepackage{epsfig}
\usepackage{color}
\usepackage{graphicx}

\def\beq{\begin{equation}}
\def\eeq{\end{equation}}

\def\be{\begin{equation}}
\def\ee{\end{equation}}
\def\bea{\begin{eqnarray}}
\def\eea{\end{eqnarray}}

\newcommand{\gsim}{\lower.7ex\hbox{$\;\stackrel{\textstyle>}{\sim}\;$}}
\newcommand{\lsim}{\lower.7ex\hbox{$\;\stackrel{\textstyle<}{\sim}\;$}}

\begin{document}

\hspace{4.5in}{}

\bigskip

\title{D0 Dimuon Asymmetry in $B_s - \bar B_s$ Mixing\\ and Constraints on New Physics}

\author{N. G. Deshpande$^{1}$, Xiao-Gang He$^{2,3}$ and German Valencia$^4$}
\affiliation{
$^1$Institute of Theoretical Science, University of Oregon, Eugene, OR 97402\\
$^2$INPAC, Department of Physics, Shanghai Jiao Tong University, Shanghai, 200240\\
$^3$Department of Physics and Center for Theoretical Sciences, National Taiwan University, Taipei\\
$^4$Department of Physics, Iowa State University, Ames, IA 50011}

\begin{abstract}

We study the consequences of the large dimuon asymmetry observed at D0. Physics beyond the standard model (SM) in $B_s-\bar B_s$ mixing is required to explain the data.
We first present a detailed analysis for model independent constraints on  physics beyond the SM, and then study the implications for theoretical models which modify the SM results in different ways, such as $Z'$ with FCNC and R-parity violating SUSY contributions.
\end{abstract}

\pacs{}

\maketitle

\noindent
\section{Introduction}

The D0 Collaboration, with 6.1 fb$^{-1}$ of data, has recently reported evidence for an anomalously large CP violation in the like-sign
dimuon charge asymmetry \cite{d0} which is attributed to semileptonic decays of $b$ hadrons defined by
\begin{eqnarray}
A_{\rm sl}^b \equiv \frac{N_b^{++}-N_b^{--}}{N_b^{++}+N_b^{--}}\; ,
\end{eqnarray}
where $N_b^{++}\;(N^{--}_b)$ is the number of events with two $b$ hadrons ($b\; \bar b$) decaying semileptonically into
$\mu^+\mu^+ X\;(\mu^- \mu^-X)$. The D0 result   \cite{d0},  $A_{\rm sl}^b = - (9.57 \pm 2.51 \pm 1.46)\times 10^{-3}$
with the first error being statistical and the second systematic,
is 3.2$\sigma$ away from the standard model (SM) prediction of $-0.2 \times 10^{-3}$   \cite{sm}.

$A_{\rm sl}^b$ is related to the asymmetries $a^{d,s}_{\rm sl}$ in $B_d$ and $B_s$ decays by
\begin{eqnarray}
A^b_{\rm sl} &=& \frac{f_d Z_d a^d_{\rm sl} + f_s Z_s a^s_{\rm sl}}{f_d Z_d + f_s Z_s}\;,
\label{As}
\end{eqnarray}
where $Z_q = 1/(1-y_q^2) - 1/(1+x^2_q)$ with $y_q = \Delta \Gamma_q/2\Gamma_q$ and $x_q = \Delta M_q/\Gamma_q$, and $f_d$ and $f_s$ stand for
the fragmentation fractions for $\bar b \to B_d$ and $\bar b \to B_s$, respectively. Using known values of  \cite{d0,PDG} $f_d = 0.323\pm 0.037$ and
$f_s = 0.118\pm 0.015$, $x_d = 0.774\pm 0.008$ and $y_d = 0$, and $x_s = 26.2\pm 0.5$ and $y_s = 0.046 \pm 0.027$, one has
\begin{eqnarray}
A^b_{\rm sl}&=& (0.506\pm 0.043) a^d_{\rm sl} + (0.494\pm 0.043)a^s_{\rm sl}\;.
\end{eqnarray}
In the above $a^q_{\rm sl}$ is the ``wrong-charge'' asymmetry,
\be
a_{\rm sl}^q \equiv \frac{\Gamma(\bar{B}_q \to \mu^+ X)-\Gamma(B_q \to \mu^-X)}{\Gamma(\bar{B}_q \to \mu^+ X)+\Gamma(B_q \to \mu^-X)} ~~.
\ee

Using the current experimental value   \cite{d0,hfag} $a_{\rm sl}^d = -0.0047\pm 0.0046$ which is consistent with zero and also with the  SM prediction  \cite{sm}
$a^d_{\rm sl} = (-4.8^{+1.0}_{-1.2})\times 10^{-4}$ within errors, one requires  \cite{d0}
\begin{eqnarray}
a^s_{\rm sl} = -0.0146\pm 0.0075\;,
\label{as-D0}
\end{eqnarray}
to obtain the D0 value of $A^b_{\rm sl}$.
This value is much larger than the SM prediction  \cite{sm} $(2.1\pm 0.6)\times 10^{-5}$ for $a^s_{\rm sl}$.

The CDF   \cite{cdf} measurement of $A_{\rm sl}^b$, using only 1.6 fb$^{-1}$ of data, has a central value which is positive,
$A_{\rm sl}^b = (8.0 \pm 9.0 \pm 6.8) \times 10^{-3} $, but is still compatible with the D0 measurement at the 1.5$\sigma$ level
because its uncertainties are 4 times larger than those of D0. Combining in quadrature (including the systematic errors)
the D0 and CDF results for $A_{\rm sl}^b$,  one finds $A_{\rm sl}^b \simeq - (8.5 \pm 2.8)\times 10^{-3}$
which is still 3$\sigma$ away from the SM value.

D0   \cite{d01} also performed a direct measurement of $a^s_{\rm sl}$,  but the result does not show any deviation from the SM, although the error bars are
quite large: $a_{\rm sl}^s = - (1.7 \pm 9.1^{+1.4}_{-1.5}) \times 10^{-3}$. This value is much smaller than the one given in Eq.~(\ref{as-D0}).
We could combine all these results  to obtain an average value $(a_{\rm sl}^s)_{\rm ave} \approx - (12.7\pm 5.0) \times 10^{-3}$.

Even though the inclusion of the CDF dimuon asymmetry and the D0 semileptonic wrong-charge asymmetry
reduces the deviation in $a_{\rm sl}^s$ derived from the D0 dimuon asymmetry, the above result
is still about 2.5$\sigma$ away from the SM value   \cite{sm}
of $(a_{\rm sl}^s)_{\rm SM}$. If confirmed, it is an indication of new physics beyond the SM \cite{su,petrov1,hv,hou,ht,petrov2,soni,dighe,fox,chen,buras,Ligeti,pich,babu,dunn}. Several theoretical attempts to explain the data have been made  \cite{dighe,fox,chen,buras,Ligeti,pich,babu,dunn}.

We note that there could be charm contamination.
It is known that there is $D^0 - \bar D^0$ mixing, and this will modify the asymmetry measured by D0 unless $D$ mesons are completely eliminated as a possible source of muons in their data sample. To take this contamination into account one would add terms proportional to $f^c_u Z^c_u$, related to charm contribution, to the formula in Eq.~(\ref{As}). Here $Z^c_u$ is analogous to $Z_i$ and is determined by the $D^0-\bar D^0$ mixing parameters $x_u$ and $y_u$, and $f^c_u$ is the fraction of direct $D^0$ and $\bar D^0$ production in $p\bar p$ collisions.  Using current values  \cite{HFAG} of $x_u = (0.98^{+0.24}_{-0.26})\%$ and $y_u = (0.83\pm 0.16)\%$ for $D^0 - \bar D^0$ mixing, the factor $Z^c_u$ can be determined. With the central values we obtain
$Z_u^c/Z_d \approx 7.2$, which is a large number. In order to obtain a reliable result, the parameter $f^c_u$ should be carefully evaluated for the muon selection criteria. With a non-zero $f^c_u$, and small CP violation in $D^0-\bar D^0$ mixing, the asymmetry will be diluted making the deviation from the SM even more severe.

\section{Constraints on new physics parameters}

The required value for $a^s_{\rm sl}$ is much larger than  the SM prediction. Attributing the observed excess to a contribution from
$B_s - \bar B_s$ mixing,  one needs to explain what type of new physics can produce the value  $(a_{\rm sl}^s)_{\rm ave} \approx - (12.7\pm 5.0) \times 10^{-3}$.
Theoretically, in terms of the mixing parameters in the $B_s-\bar B_s$ system, to a very good approximation $a^s_{\rm sl}$ is given by  \cite{sm}
\begin{eqnarray}
a^s_{\rm sl} =  \frac{|\Gamma^{12}_s|}{|M_s^{12, SM}|}\frac{\sin\phi_s}{|\Delta_s|} = (4.97\pm 0.94)\times 10^{-3}\frac{\sin\phi_s}{|\Delta_s|}\;.
\label{as-f}
\end{eqnarray}
where $\Delta_s$ and the phase $\phi_s$ are defined by, $M_s^{12} = M_s^{12,SM} + M^{12,NP}_s = M_s^{12,SM}\Delta_s = |M^{12,SM}_s| |\Delta_s|e^{i\phi_s}$. We adopt the phase convention that $\Gamma^{12}_s$ is real.

Since the SM prediction for $\Delta M_s \approx 2 |M^{12}_s|$ agrees with data well, $|\Delta_s|$ is only allowed to vary in a limited region, $0.92\pm 0.32$ fixed by the experimental value  \cite{PDG}
$\Delta M_s = (17.77\pm 0.12)ps^{-1}$ and the SM prediction  \cite{sm} $(19.30\pm 6.74)ps^{-1}$. In order to reproduce the  D0
result, it would seem naively that $\sin\phi_s$ would have to exceed the physical range  $|\sin \phi_s| < 1$,   as one needs $\sin\phi_s = -2.56\pm 1.16$. This situation calls for a more careful analysis  considering $\phi_s$ and $|\Delta_s|$ simultaneously. Here we wish to point out that Eq.~(\ref{as-f}) is only an approximation, and we now review the derivation of the exact formula for the asymmetry.

Denoting the element in the Hamiltonian responsible for  $B_s - \bar B_s$ mixing by $M^{12}_s - i \Gamma^{12}_s/2$, and working in a basis where $-\Gamma^{12}_s = |\Gamma^{12}_s|$ is positive real, one can write the same element as $|M^{12}_s|e^{i\phi_s} + i |\Gamma^{12}_s|/2$ (this is equivalent to defining $\phi_s = arg(-M^{12}_s/\Gamma^{12}_s)$). We have
\begin{eqnarray}
&&\Delta M_s + i\Delta \Gamma_s/2 = 2\sqrt{(M_s^{12} - i \Gamma^{12}/2)(M^{12*}_s-i\Gamma^{12*}/2)}\;,\nonumber\\
&&\Delta M_s \Delta \Gamma_s = 4|M^{12}_s||\Gamma^{12}_s|\cos\phi_s\,\;,\;\;\Delta M_s^2 - \Delta \Gamma^2_s/4 = 4(|M^{12}_s|^2-|\Gamma^{12}_s|/4)\;.
\end{eqnarray}
Note that the above definitions of $\Delta M_s$ and $\Delta \Gamma_s$ are the same as those in Ref.  \cite{d0}.

Further defining
$w_s=2|M^{12}_s|/\Gamma_s$, and $z_s = |\Gamma^{12}_s|/\Gamma_s$, we have
\begin{eqnarray}
&&w_s^2-z_s^2 = x_s^2-y_s^2\;,\;\;w_sz_s\cos\phi_s = x_sy_s\;,
\end{eqnarray}
which lead to
\begin{eqnarray}
&&w_s^2 = \frac{1}{2}\left(x_s^2-y_s^2 + \sqrt{(x_s^2-y_s^2)^2 + \frac{4x_s^2y_s^2}{\cos^2\phi_s}}\right)\;,\nonumber\\
&&z_s^2 = \frac{1}{2}\left(y_s^2-x_s^2 + \sqrt{(x_s^2-y_s^2)^2 + \frac{4x_s^2y_s^2}{\cos^2\phi_s}}\right)\;.
\end{eqnarray}

The asymmetry $a^s_{\rm sl}$ in terms of $w_s$, $z_s$ and $\sin\phi_s$ is given by  \cite{hv}
\begin{eqnarray}
a^s_{\rm sl} &=& \frac{2 w_s z_s \sin\phi_s}{w_s^2+z_s^2} = \frac{\sin\phi_s}{\sqrt{1+\frac{(x_s^2-y_s^2)^2}{4x_s^2y_s^2}\cos^2\phi_s}}\nonumber\\
&=&\frac{\sin\phi_s}{\sqrt{1+\frac{(1-(\Delta\Gamma_s/2\Delta M_s)^2)^2}{4(\Delta \Gamma_s/2\Delta M_s)^2}\cos^2\phi_s}}\;.
\label{exact-a}
\end{eqnarray}

The fact that $w_s>>z_s$ allows one to write an approximate formula
\begin{eqnarray}
a^s_{\rm sl} \approx \frac{2z_s}{w_s}\sin\phi_s = \frac{|\Gamma^{12}_s|}{|M^{12}_s|}\sin\phi_s\;.
\end{eqnarray}
This is the  formula that is often used and  given in Eq.~(\ref{as-f}).

A careful expansion of $a^s_{\rm sl}$ in terms of $x_s$ and $y_s$, reveals that
\begin{eqnarray}
a^s_{\rm sl} = \left \{ \begin{array}{ll}
\frac{\Delta \Gamma_s}{\Delta M_s} \tan\phi_s&\;\;\;\;\;\frac{\Delta \Gamma_s}{\Delta M_s}<<\cos\phi_s\\
\sin\phi_s &\;\;\;\;\;\frac{\Delta \Gamma_s}{\Delta M_s}>>\cos\phi_s\\
\end{array}
\right .
\end{eqnarray}
Note that the asymmetry can be as large as order one for fixed $\Delta M_s$ and $\Delta \Gamma_s$.

In the SM, the phase in $M_s^{12,SM}$ is  \cite{sm} $0.0041\pm 0.0014$ which is too small to play a substantial role in explaining the large asymmetry observed. New physics beyond the
SM may induce large CP violating phases and also change the magnitudes for both $M_s^{12}$ and $\Gamma^{12}_s$.

We now consider the constraints on the
new physics contribution to $M_s^{12,NP} = |M_s^{12,NP}|e^{i\phi_{NP}}$ assuming that there is no alteration to the SM prediction for $\Gamma^{12}_s$. Defining $R=|M^{12,NP}_s|/|M^{12,SM}_s|$ and neglecting the small phase in $M_s^{12,SM}$, we can solve for $\sin\phi_s$ and $\Delta_s$,
\begin{eqnarray}
&&\frac{\sin\phi_s}{|\Delta_s|} = \frac{R \sin\phi_{NP}}{(1+2R\cos\phi_{NP} + R^2)}\;,\nonumber\\
&&|\Delta_s| = (1+2 R\cos\phi_{NP} + R^2)^{1/2}
\end{eqnarray}
Whether there are physical solutions for $R$ and $\phi_{NP}$, should be analyzed using the above equations with the constraints for $\sin\phi_{s}/|\Delta_{s}|$ and $|\Delta_s|$ in the ranges $(-2.56\pm 1.16)$ and $(0.92\pm 0.32)$, respectively.

In Fig.1 we show the ranges for $|\Delta_s|$ and $R$ for a given  value of $\delta \equiv \sin\phi_s/|\Delta_s| = a^s_{\rm sl}/4.97\times 10^{-3}$ (or $a^s_{\rm sl} = 4.97\times 10^{-3}\delta$) as a function of $\sin \phi_{NP}$. Since D0 data requires that the asymmetry $a^s_{\rm sl}$ be negative, this restricts $\sin\phi_{NP}$ to be negative too. One  can then see in which quadrant  should $\phi_{NP}$ be in order to reproduce the data. To obtain a large size for $\delta$, a lower value of $|\Delta_s|$ is preferred, therefore the solution with $\cos\phi_s <0$  is preferred.

From the figures, we see that it is not possible for $\delta$ to get down to the central value $-2.56$ as this would require a value for $|\Delta_s|$ below its one sigma lower bound. We have checked that in order to have solution within the one $\sigma$ region of $|\Delta_s|$, $\delta$ can at most go down to -1.6. To have $\delta$ reach the D0 central value  it is also necessary to modify $\Gamma^{12}_s$.
The analysis, in general, will now be different  \cite{hv}. However, since
$|M^{12}_s|$ is much larger than $|\Gamma^{12}_s|$, the change of $\Gamma^{12}_s$ needed can be easily accommodated in Eq.~(\ref{as-f}) by multiplying by a factor $\theta = |\Gamma^{12}_s|/|\Gamma^{12,SM}_{12}|$ and modifying the phase $\phi_s$ to include the contribution from $\Gamma^{12}_s$. The value for $a^s_{\rm sl}$ is then scaled by a factor $\theta$.
The central value of the D0 asymmetry is then obtained with $\theta$ around 1.6, which is allowed by the experimental data on $\Delta \Gamma_s$.
It should be noted that the usually quoted value for $\Gamma^{12}_s$ is from short distance SM contribution \cite{sm}, there may be long distance contributions which  modify the value. A larger than SM short distance contribution to $\Gamma^{12}_s$ is still a possibility. However, it is difficult to reliably calculate the long distance contribution. It is also possible that a large $\Gamma^{12}_s$ is due to new physics beyond the SM \cite{petrov1,ht,dighe}.

\begin{figure}[hb]
\includegraphics[width=1.5in]{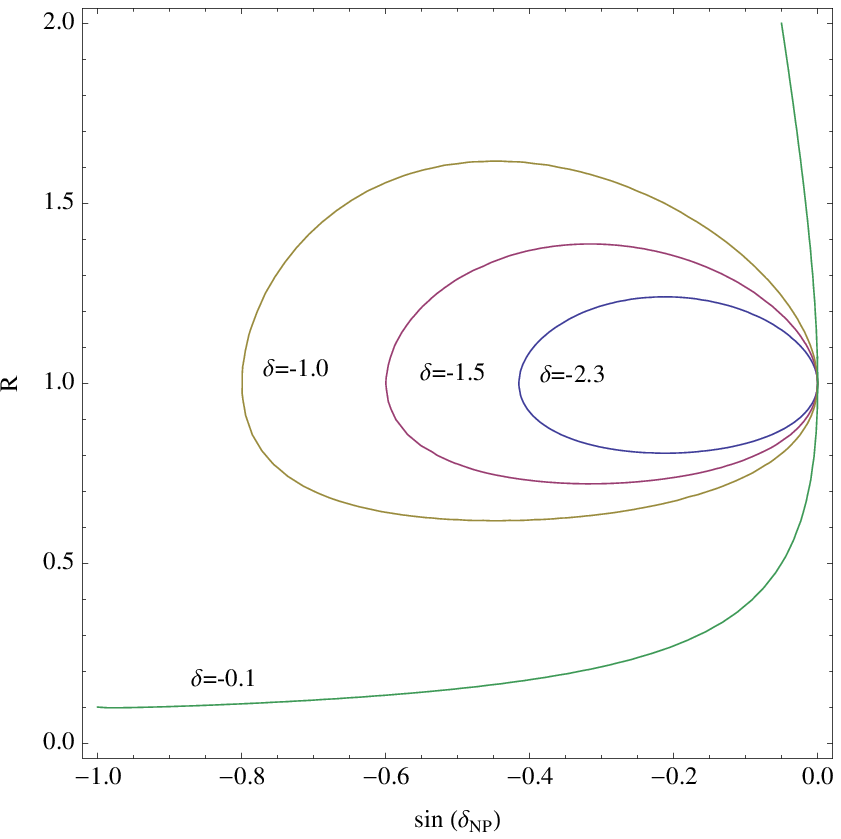}
\hspace{0.5cm}\includegraphics[width=2.3in]{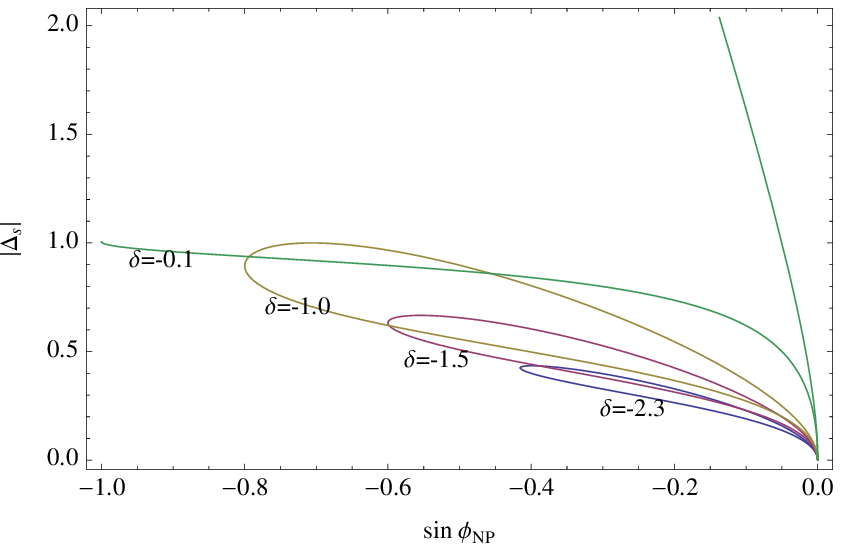}
\caption{$R$ and $|\Delta_s|$ as functions of $\sin\phi_{NP}$ for
representative values of $\delta$.  All curves use the central value
for $M^{12,SM}$.
\label{fig1}}
\end{figure}

\begin{figure}[hb]
\includegraphics[width=2.0in]{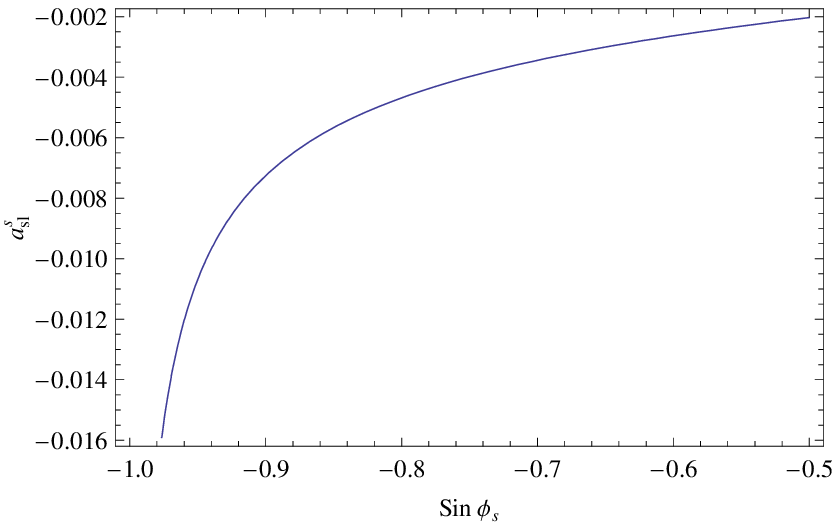}
\hspace{0.3cm}\includegraphics[width=2.0in]{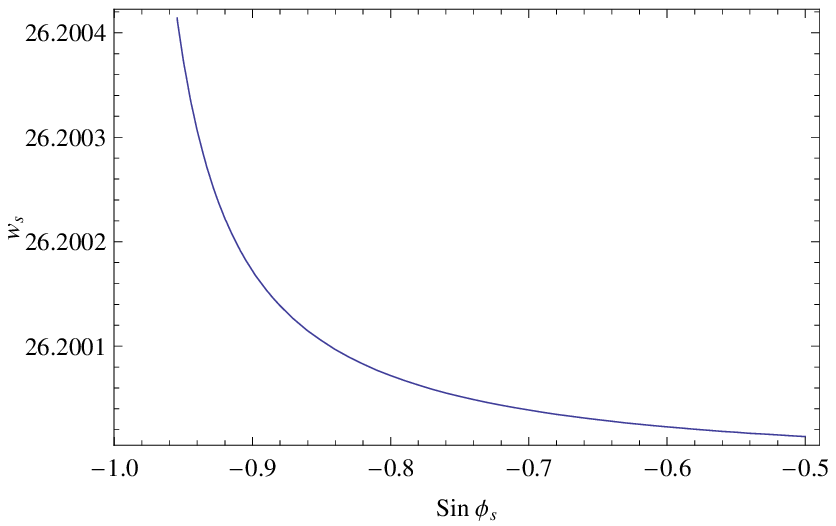}
\hspace{0.3cm}\includegraphics[width=1.9in]{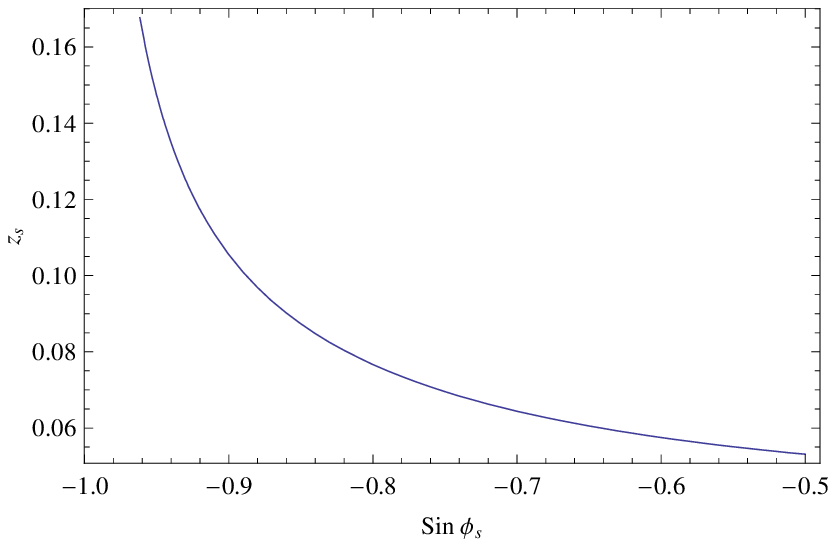}
\caption{From left to right $a^s_{\rm sl}$, $w_s = 2|M^{12}_s|/\Gamma_s$ and $z_s = |\Gamma^{12}_s|/\Gamma_s$ as functions of $\sin\phi_s$ respectively. In all cases we use the experimental central values $x_s =26.2$ and $y_s = 0.046$.}
\label{exp}
\end{figure}

When we go beyond the SM, new contributions are in general not known. It is therefore desirable to use experimentally measurable quantities as much as possible.
The exact formula in Eq.~(\ref{exact-a}) allows one to predict  $a^s_{\rm sl}$ using the measured values $x_s$ and $y_s$, and a theoretically unknown phase $\phi_s$. Taking the experimental values  \cite{PDG}
$x_s = 26.2\pm 0.5$ and $y_s = 0.046\pm 0.027$, one can ask what theoretical values for $|M^{12}_s|$, $|\Gamma^{12}_s|$ and $\sin\phi_s$ are needed. We show the results in Fig.\ref{exp}.

The central value of dimuon asymmetry can be reproduced with $\sin\phi_s$ around - 0.96, and $z_s$ around 0.16. This implies that $\Gamma^{12}_s$ is a factor of 3 larger than the usual SM short distance contribution.

\section{Implications for Beyond SM model parameters}

In order to reproduce the anomalously large dimuon asymmetry observed by D0, new physics beyond the SM is needed. There are different ways in which the dimuon asymmetry in $B_s - \bar B_s$ can be affected.  For example
$Z^\prime$ models with tree level FCNC  \cite{zpfcnc1,zpfcnc2,hv} can easily induce a large modification for $M^{12}$, but have little effect on $\Gamma^{12}_s$.  R-parity
violating interactions in SUSY models can also have tree level FCNC. Besides modifying $M^{12}$, they can  induce modifications in $\Gamma^{12}_s$ \cite{ht}. However, taking into account new constraints from various experimental data \cite{hv,ht,zpfcnc1,zpfcnc2,german,r-constraints}, it is difficult to generate sizeable modifications in $\Gamma^{12}$~\footnote{We than C. Bauer and N. Dunn for bringing these new constraints to our attention.}. In the following we discuss these two types of models and their
contributions to $M^{12}_s$ and $\Gamma^{12}_s$, taking into account constraints from other possible data,  in some detail.

\subsection{$Z'$ model with FCNC}

In general a $Z'$ can couple to quarks with FCNC of the form  \cite{zpfcnc2}
\begin{eqnarray}
{\cal L} = {g\over 2 c_W} \bar q_i \gamma^\mu (a_{ij}P_L +
b_{ij}P_R )q_j Z^\prime_\mu\;.
\end{eqnarray}
In $Z'$ models, the new contributions to $M^{12}_s$ happen at tree-level, whereas the modification of $\Gamma^{12}_s$  only occurs at the loop level and is therefore a smaller effect.  We will concentrate on the tree level $Z'$ exchange contribution to $M^{12}_s$ with constraints on the parameters from other processes \cite{zpfcnc1,zpfcnc2,german}.
We begin from the known contribution to $M_{12}$ from $Z^\prime$ exchange  \cite{zpfcnc2},
\begin{eqnarray}
M^{12,Z^\prime}_s &=& {G_F\over \sqrt{2}}{m^2_Z\over
m^2_{Z^\prime}}\eta_{Z^\prime}^{6/23}{1\over 3} f^2_{B_s} M_{B_s} B_{B_s}
( a^2_{sb}
+ b^2_{sb} \nonumber\\
& +& \eta_{Z^\prime}^{-3/23} {1\over 2}
a_{sb}b_{sb}(2\epsilon -3) + {2\over
3}(\eta_{Z^\prime}^{-3/23}-\eta_{Z^\prime}^{-30/23}){1\over
4}a_{sb}b_{sb} (1-6\epsilon)).
\end{eqnarray}
where  $\eta_{Z^\prime} \equiv \alpha_s(m_{Z'})/\alpha_s(\mu)$ is a  QCD correction resulting from the running scale from $m_{Z'}$ to $\mu =m_b$.
$B_{B_s}$ is the ratio between the matrix
element  $<\bar B_s|\bar s\gamma^\mu \gamma_5 b \bar s \gamma_\mu
\gamma_5 b|\bar B_s>$ and its value in factorization. Similarly,
$\epsilon$ is defined as $\epsilon = (\tilde
B_{LR}/B_{B_s})(m^2_{B_s}/(m_s+m_b)^2)$ where $\tilde B_{LR}$ is the
ratio between the matrix element $<\bar B_s|\bar s \gamma_5 b \bar s
\gamma_5 b|\bar B_s>$ and its value in factorization. When needed, we will use
$\epsilon =1$ in our numerical results.

Using the central value from lattice calculation \cite{lattice} $f_{B_s}\sqrt{B_{B_s}} = 270$ MeV, we obtain
\begin{eqnarray}
\Delta_s &=& 1 + 6.8\times 10^4 \left (\frac{m_Z}{m_{Z'}}\right )^2
\nonumber\\
&\times &\left ( a^2_{sb}
+ b^2_{sb} + \eta_{Z^\prime}^{-3/23} {1\over 2}
a_{sb}b_{sb}(2\epsilon -3) + {2\over
3}(\eta_{Z^\prime}^{-3/23}-\eta_{Z^\prime}^{-30/23}){1\over
4}a_{sb}b_{sb} (1-6\epsilon)\right ).
\end{eqnarray}

To make sure that there are solutions for the required value of $a^s_{\rm sl}$ determined from D0 data,
we take a case with $b_{sb} = 0$ for illustration. In this case we have
\begin{eqnarray}
R = 6.8\times 10^4 \left (\frac{m_Z}{m_{Z'}}\right )^2 |a_{sb}|^2,\;\;
\phi_{NP} = 2arg(a_{sb}) \;.
\label{xy}
\end{eqnarray}

As we have discussed earlier, within the one $\sigma$ allowed region for $|\Delta_s|$ it is not possible to obtain the D0 central value for $a^s_{\rm sl}$. To illustrate the range of $a^s_{\rm sl}$ that can be obtained with this model, we consider a few specific values of $\delta = \sin\phi_s/|\Delta_s|$ and restrict $|\Delta_s|$  to be within its one $\sigma$ allowed region.

A solution with an asymmetry within one $\sigma$ of the D0 result requires $\delta$ to be less than -1.4. For illustration we take $\delta = -1.5$ (corresponding to  $a^s_{\rm sl}$ in the range of $(-0.78 \sim -6.0)\times 10^{-3}$). With $\delta = - 1.5$, $\sin\phi_{NP}$ is restricted to be in the range $-0.39 \sim -0.59$ and the corresponding range for $R$ is $1.37 \sim 0.91$, as can be seen in Figure~\ref{fig1}. If the large asymmetry observed by D0 is confirmed, the new physics parameters $R$ and $\sin\phi_{NP}$ need to be in the above ranges. However, if it turns out that the asymmetry is smaller, one needs to use a smaller $\delta$, in which case solutions are much easier to obtain. We will concentrate on the case with $\delta  = -1.5$. Applying  the above ranges of parameters to Eq.~(\ref{xy}), we then have
\begin{eqnarray}
&&arg(a_{sb}) \mbox{ is in the range:}\;\; (\frac{\pi}{2}+ 0.19 \sim \frac{\pi}{2}+0.31)\;,\nonumber\\
&&\frac{m_Z}{m_Z'}|a_{sb}| \mbox{ is in the range:} \;\;(0.0036 \sim 0.0044)\;.
\end{eqnarray}
This range is comparable to other constrains on FCNC from non-universal $Z^\prime$ models \cite{zpfcnc1}, and in particular admits solutions with small $|a_{sb}|$, say of $O(10^{-2})$.
We note that $Z^\prime$ couplings to $b\bar s$ quarks has very weak constraints from the decays $B_s \to \mu^+\mu^-$ and $B_d \to K l^+l^-$ which are much lower than values used above.
In this case the $Z'$ mass can be in the several hundreds of GeV, a region that can be studied at the LHC. We note that models with natural suppression of flavor changing couplings exist in the literature  \cite{german}.

The case with $a_{sb} = 0$ is the same as the case discussed above, but with $a_{sb}$ replaced by $b_{sb}$. If both $a_{sb}$ and $b_{sb}$ are non-zero, the analysis is more complicated. For example, for the special case with $a_{sb} = b_{sb}$ the contribution from $Z'$ exchange to $R$ is reduced by a factor of 0.6. This translates into the coupling, $a_{sb} = b_{sb}$, being enhanced by a factor $1.3$ and the phase range being the same as in the $b_{sb}=0$ case.

\subsection{SUSY models with R-parity violation}

We now discuss an example which can modify $\Gamma^{12}_s$, an R-parity violating (RPV) interaction in SUSY models. However, we find that existing constraints \cite{r-constraints} will limit the effect to levels below what is required to explain the D0 data.

There are three types of $R$-Parity violating terms
  \cite{rbreaking}:
${\lambda_{ijk}/ 2} L_L^iL_L^j  E_R^{ck}$, $\lambda'_{ijk}
L_L^iQ_L^j D_R^{ck}$, and ${\lambda''_{ijk}/ 2} U_R^{ci}
D_R^{cj} D_R^{ck}$.
Here $i,j$ and $k$ are the generation indices: $L_L^{}, Q_L^{},
E_R^{}, D_R^{}$ and $U_R^{}$ are the left handed lepton, quark, right handed
lepton, down-quark and up-quark fields, respective.
%
$\psi^c$ indicates the charge conjugated field of $\psi$
The contributions of these interactions to $\Delta
\Gamma^{12}_s$ have been studied in detail  \cite{ht}. It is found that
these interactions can induce a non-zero $M^{12}_s$ at tree level. There are also couplings that can induce a non-zero $\Gamma^{12}_s$ at one loop level without tree level contributions to $M_s^{12}$.

The contributions to $z_s$ can be grouped into several categories according to particles exchanged \cite{ht}. For $\lambda'$ couplings, these contributions are:
$z_s(SM - RPV)$ from interference between SM and R-parity violating interactions; $z_s(RPV - RPV,\nu)$ from exchanges of neutrinos and down-type squarks; $z_s(RPV - RPV,l)$ from exchanges of
charged leptons and up-type squarks; $z_s(RPV - RPV,u)$ from exchanges of light up-type  quarks and charged  sleptons; and $z_s(RPV - RPV, d)$ from exchanges of light down-type quarks and sneutrinos. For $\lambda^{\prime\prime}$ couplings, the contributions are: $z_s(SM - RPV)$ from interference between SM and R-parity violating interactions; $z_s(RPV - RPV, u)$ from exchanges of
up-type  light quarks and down-type squarks; and $z_s(RPV - RPV, d)$ from exchanges of
down-type  light quarks and up-type  squarks.

Although the couplings involved do not contribute to $B_s - \bar B_s$ mixing at tree level, they contribute to various $B$ decays, such as $b \to s \gamma$ and $B\to MM$ (where $M$ is a light pseudoscalar meson). After applying these constraints, the most likely large contributions are given by
\begin{eqnarray}\label{eq:Bs1}
&&z_s(SM - RPV) = -13 (\lambda'_{i22}\lambda^{\prime*}_{i23} + \lambda'_{i22}\lambda^{\prime *}_{i13})\frac{(100GeV)^2}{m^2_{\tilde {e}^{i}_L}}\;,\nonumber\\
&&z_s(RPV - RPV,d) = - 233
\times 28\times \lambda'_{ijj'}
\lambda'^*_{i23}\lambda'_{i'32}\lambda'^*_{i'jj'}
\frac{(100GeV)^4}{m^2_{\tilde {\nu}^i_L} m^2_{\tilde {\nu}^{i'}_L}}
\;,\\
&&z_s(SM - RPV) = -2.9\lambda^{\prime\prime}_{221}\lambda^{\prime\prime *}_{231}\frac{(100GeV)^2}{m^2_{\tilde {d}^{1}_L}}\;,\nonumber
\end{eqnarray}

The constraints on these couplings  that arise from $B$ decays, assuming sparticles with mass 100 GeV, are \cite{r-constraints}, $|\lambda^\prime_{i32}\lambda^{\prime *}_{i22}| \sim |\lambda^\prime_{i23}\lambda^{\prime *}_{i22}| < 2.3\times 10^{-3}$,
$|\lambda^\prime_{i13}\lambda^{\prime *}_{i22}| < 2.48\times 10^{-3}$, and $|\lambda^{\prime\prime}_{221}\lambda^{\prime\prime *}_{231}| < 10^{-2}$.  These numbers constrain the three contributions in Eq.~\ref{eq:Bs1} to be less than 0.06, 0.03 and 0.03, respectively. All of them are smaller by factors of 3-5 than the $z_s \sim 0.16$ required to explain the D0 asymmetry.
We cannot rule out the possibility that all these contributions (plus the SM) add up constructively to reach the value required  by D0 data, but at this stage it is fair to conclude that this is not a favored explanation.

\section{Time dependent CP violation with a non-zero $\Delta \Gamma$}

Finally, we comment on the CP asymmetry
$A_{TCP}$ which can be measured by studying the  time dependent $B \to l^+ \nu
X$ and $\bar B \to l^- \bar \nu \bar X$ decay rate difference,
\begin{eqnarray}
&&A_{TCP} = 2 e^{\frac{\Delta \Gamma_s}{2} t} {A_f \cos(\Delta
M_s t) + S_f \sin(\Delta M_s t)\over 1+ e^{\Delta
\Gamma_s t} - A^{\Delta \Gamma}_f (1- e^{\Delta \Gamma_s t})},
\end{eqnarray}
where $f$ is not a CP eigenstate. The notation follows Ref.~\cite{zpfcnc2},
\begin{eqnarray}
&&A_f = {|A(f)|^2-|\bar A(\bar f)|^2\over |A(f)|^2+|\bar A(\bar
f)|^2},\;\;S_f = - 2 {Im((q_{B_s}/p_{B_s})\bar A(f)A^*(f))\over
|A(f)|^2+|\bar A(\bar f)|^2},\nonumber\\
&&A^{\Delta \Gamma}_f =  2 {Re((q_{B_s}/p_{B_s})\bar
A(f)A^*(f))\over |A(f)|^2+|\bar A(\bar f)|^2},
\;\;|A_f|^2+|S_f|^2+|A^{\Delta \Gamma}_f|^2 = 1.
 \label{qm}
\end{eqnarray}
In this equation, $A(f)$ and $\bar A(\bar f)$ are time dependent decay amplitudes for $B_s$
and $\bar B_s$ decay into states $f$ and $\bar f$ in terms of the $B_s$ mixing parameters
\begin{eqnarray}
{q_{B_s}\over p_{B_s}} = \sqrt{{M_s^{12*} -
i\Gamma_s^{12*}/2\over M_s^{12} - i \Gamma_s^{12}/2}}.
\end{eqnarray}

Assuming that CP violation in $A$ and $\bar A$ is small, $|A| =
|\bar A|$, one obtains
\begin{eqnarray}
A_{TCP} = 2 e^{\frac{\Delta \Gamma_s t}{2}} {\sin\phi_s \sin(\Delta
M_s t)\over 1+ e^{\Delta \Gamma_s t} - \cos\phi_s (1-
e^{\Delta \Gamma_s t})}.
\end{eqnarray}

\begin{figure}[hb]
\includegraphics[width=2.5in]{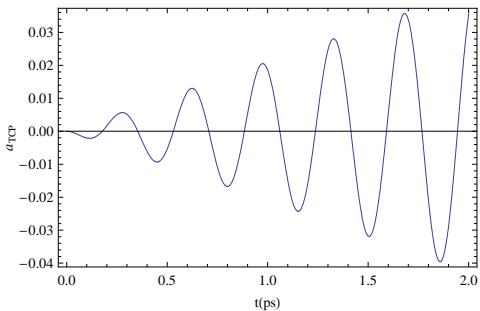}
\caption{$a_{TCP}$ vs. t(ps) with $\sin \phi_s = -0.5$ and $\cos\phi_s <0$.
\label{quantf}}
\end{figure}

If $\Delta \Gamma_q=0$, which is a very good
approximation for $B_d$ decays, it is not possible to study the quantity
$A^{\Delta \Gamma}_f$ in totality. The time dependence is a simple sine function of time
and one cannot check the $\Delta \Gamma$ effect.
Therefore this CP violating observable is special to $B_s -\bar B_s$ system
because $\Delta \Gamma_s$ is not zero. It offers a possibility
to study the exponential factor in the time dependence.
In the SM the phase $\phi_s$ is small  \cite{sm}, $0.0041\pm 0.0014$, resulting in
a very small $A_{TCP}$. If the D0 result is confirmed however, a large phase is
allowed as we saw earlier and a large $A_{TCP}$ is possible. In Fig(\ref{quantf}),
we show $ a_{TCP}=A_{TCP}(\Delta
\Gamma_s) - A_{TCP}(0)$ as a function of t. We have chosen the value $\sin\phi_s =-0.5$ with
the phase in the third quadrant and the central value  $\Delta_s=0.096 $ for illustration.
We can see that at the few percent level, there are differences
with respect to the $\Delta \Gamma_s =0$ case, and such differences  may be tested at the LHCb or at a B-factory, such as BELLE II with sufficient high
energy to produce $B_s \bar B_s$ pairs.

\section{Summary}

The dimuon asymmetry reported by D0 is much larger than the SM prediction although further experimental studies are needed to confirm this  result.  $D^0 - \bar D^0$ mixing  may contaminate the final results and it is therefore important to carry out a detailed study with careful selection criteria for the dimuon events. If the D0 result is confirmed, it represents clear evidence for new physics beyond the SM.

If new physics only affects $M^{12}_s$, we can get an asymmetry within one $\sigma$ of the D0 value, but it is not possible to reach the central value. Modification of $\Gamma^{12}_s$ by new physics may then play an important role. We studied the consequences of the D0 dimuon asymmetry on a $Z^\prime$ model with tree level FCNC and in a SUSY model with R-parity violating interactions. We find that the exchange of $Z'$  can significantly modify $M^{12}_s$ and bring the theoretical prediction to within one $\sigma$ of the D0 allowed region. We showed that there are R-parity violating terms which can modify $\Gamma^{12}_s$ but that these modifications are probably too small to account for the observed D0 asymmetry. We also showed that the
D0 result implies a large effect on time dependent CP violation in $B_s -\bar B_s$ mixing resulting in a detectable non-zero $\Delta \Gamma_s$ effect at the LHCb.

\bigskip

\newpage
\noindent
{\bf Acknowledgements}


\noindent
We thank  G. Brooijmans, W.-S. Hou and D. Soper for discussions.
This work is supported in party by DOE under contract numbers DE-FG02-96ER40969ER41155 and DE-FG02-01ER41155, and in part by the NSC and NCTS
of ROC. XGH thanks the Institute of Theoretical Science at University of Oregon for hospitality where part of this work was done.


\bigskip
\bigskip


\begin{thebibliography}{99}
\bibitem{d0} V.M. Abazov et al. [D0 Collaboration], arXiv:1005.2757[hep-ex].
\bibitem{sm} A. Lenz and U. Nierste, J. High Energy Phys. {\bf 0706}, 072(2007)

\bibitem{PDG} Particle Data Group, C. Amsler, et al., Phys. Lett. {\bf B667}, 1 (2008), and 2010 edition.
\bibitem{hfag} E. Barberio et al. (HFAG), arXiv:0808.1297[hep-ex].

\bibitem{cdf} CDF Collaboration, Note 9015, Oct. 2007.
\bibitem{d01} V.M. Abzov et al. [D0 Collaboration], arXiv:0904.3907[hep-ex].



\bibitem{su} L. Randall and S.-F. Su, Nucl. Phys. {\bf B540}, 37(1999).




\bibitem{petrov1} A. Badin, F. Gabbiani and A. Petrov, Phys. Lett. {\bf B653}, 230(2007).

 \bibitem{hv}
  X.~G.~He and G.~Valencia,
  Phys.\ Lett.\  B {\bf 651}, 135 (2007)
  [arXiv:hep-ph/0703270].


\bibitem{hou} W.~S.~Hou and N.~Mahajan,
  Phys.\ Rev.\  D {\bf 75}, 077501 (2007)
  [arXiv:hep-ph/0702163].

\bibitem{ht} Shao-Long Chen, Xiao-Gang He, A. Hovhannisyan, Ho-Chin Tsai, J. High Energy Phys. {\bf 0709}, 044,2007.

\bibitem{petrov2} A. Petrov and G. Yeghiyan, Phys. Rev. {\bf D77}, 034018(2008).

\bibitem{soni} A.~Soni, A.~K.~Alok, A.~Giri, R.~Mohanta and S.~Nandi,
  arXiv:1002.0595 [hep-ph].


\bibitem{dighe} Amol Dighe, Anirban Kundu, and Soumitra Nandi,
e-Print: arXiv:1005.4051 [hep-ph].
\bibitem{fox} Bogdan A. Dobrescu, Patrick J. Fox, Adam Martin,
e-Print: arXiv:1005.4238 [hep-ph].
\bibitem{chen} Chuan-Hung Chen, Gaber Faisel,
e-Print: arXiv:1005.4582 [hep-ph].
\bibitem{buras} A. Buras, S. Gori, M. Carlucci and G. Isidori, e-print: arXiv:1005.5310[hep-ph].


\bibitem{Ligeti}
  Z.~Ligeti, M.~Papucci, G.~Perez and J.~Zupan,
  arXiv:1006.0432 [hep-ph].
\bibitem{pich} M. Jung, A. Pich and P. Tuzon, e-print: arXiv:1006.0470[hep-ph].

\bibitem{babu}
K.~S.~Babu and J.~Julio,
  arXiv:1006.1092 [hep-ph].

\bibitem{dunn} C. Bauer and N. Dunn, e-print: arXiv:1006.1629[hep-ph]

\bibitem{HFAG} Heavy Flavor Averaging Group: http://www.slac.stanford.edu/xorg/hfag/charm/index.html



\bibitem{zpfcnc1}
  X.~G.~He and G.~Valencia,
  Phys.\ Rev.\  D {\bf 66}, 013004 (2002)
  [Erratum-ibid.\  D {\bf 66}, 079901 (2002)]
  [arXiv:hep-ph/0203036];
  X.~G.~M.~He and G.~Valencia,
  Phys.\ Rev.\  D {\bf 70}, 053003 (2004)
  [arXiv:hep-ph/0404229];
K. Cheung, C.-W. Cheng, N. Deshpande and J. Jiang, Phys.Lett.{\bf B652}, 285-291(2007)
C.-W. Cheng, N. Deshpande and J. Jiang, J. High Energy Phys. {\bf 0608}, 075(2006); V. Barger, C.W. Chiang, J. Jiang and
P. Langacker, Phys. Lett. {\bf B596}, 229(2004); V. Barger et al., J. High Energy Phys. {\bf 0912}, 048(2009); Phys. Rev. {\bf D80}, 055008(2009).
\bibitem{zpfcnc2}
  X.~G.~He and G.~Valencia,
  Phys.\ Rev.\  D {\bf 74}, 013011 (2006)
  [arXiv:hep-ph/0605202].



\bibitem{german} C.~W.~Chiang, N.~G.~Deshpande and J.~Jiang,
  JHEP {\bf 0608}, 075 (2006)
  [arXiv:hep-ph/0606122];
P. Langacker, Rev. Mod. Phys. {\bf 81}, 1199(2008)[arXiv:0801.1345 [hep-ph]];
  X.~G.~He and G.~Valencia,
  Phys.\ Lett.\  B {\bf 680}, 72 (2009)
  [arXiv:0907.4034 [hep-ph]];
C.~W.~Chiang, N.~G.~Deshpande, X.~G.~He and J.~Jiang,
  Phys.\ Rev.\  D {\bf 81}, 015006 (2010)
  [arXiv:0911.1480 [hep-ph]].






\bibitem{r-constraints}
 M.~Chemtob, Prog.\ Part.\ Nucl.\ Phys\. {\bf 54}, 71 (2005)
  [arXiv:hep-ph/0406029];
  A.~Kundu and J.~P.~Saha,
  Phys.\ Rev.\  D {\bf 70}, 096002 (2004)
  [arXiv:hep-ph/0403154];
  S.~Nandi and J.~P.~Saha,
  Phys.\ Rev.\  D {\bf 74}, 095007 (2006)
  [arXiv:hep-ph/0608341];
  D.~K.~Ghosh, X.~G.~He, B.~H.~J.~McKellar and J.~Q.~J.~Shi,
  JHEP {\bf 0207}, 067 (2002)
  [arXiv:hep-ph/0111106].

\bibitem{lattice} V. Lubicz and C. Tarantino, e-print arXiv:0807.4605.

\bibitem{rbreaking} C. Aulak and R. Mohapatra, 
Phys. Lett. {\bf B119}, 136 (1983);
F. Zwirner, 
Phys. Lett. {\bf B132}, 103 (1983);
L. J. Hall and M. Suzuki, 
Nucl. Phys. {\bf B231}, 419 (1984);
I. H. Lee, 
Nucl. Phys. {\bf B246}, 120 (1984);
J. Ellis {\it et al.}, 
Phys. Lett. {\bf B150}, 142 (1985);
G. G. Ross and J. W. F. Valle, 
Phys. Lett. {\bf B151}, 375 (1985);
S. Dawson, 
Nucl. Phys. {\bf B261}, 297 (1985);
R. Barbieri and A. Masiero, 
Nucl. Phys. {\bf B267}, 679 (1986).

\end{thebibliography}
\end{document}